\documentclass[useAMS,macros,usenatbib,usegraphicx,a4paper]{mn2e}
\pdfoutput=1

\usepackage[usenames,dvipsnames]{color}
\usepackage{amsfonts}
\usepackage{multirow}
\usepackage{graphicx}
\usepackage{amsmath}
\usepackage[xindy]{glossaries}
\usepackage{aas_macros}
\usepackage{subfigure}
\usepackage{rotating}
\usepackage{hyperref}

\makeglossaries

\newacronym{10gbe}{10 GbE}{10 Gigabit Ethernet}
\newacronym{adc}{ADC}{Analogue to Digital Converter}
\newacronym{best2}{BEST-2}{Basic Element for SKA Training II}
\newacronym{cmac}{CMAC}{Complex Multiply and Accumulate}
\newacronym{casper}{CASPER}{Collaboration for Astronomy Signal Processing and Electronics Research}
\newacronym{dsp}{DSP}{Digital Signal Processing}
\newacronym{eor}{EoR}{Epoch of Reionization}
\newacronym{fft}{FFT}{fast Fourier transform}
\newacronym{fpga}{FPGA}{Field Programmable Gate Array}
\newacronym{gpu}{GPU}{Graphics Processing Unit}
\newacronym{hdl}{HDL}{Hardware Description Language}
\newacronym{if}{IF}{Intermediate Frequency}
\newacronym{lo}{LO}{Local Oscillator}
\newacronym{lvds}{LVDS}{Low-Voltage Differential Signal}
\newacronym{pfb}{PFB}{Polyphase Filterbank}
\newacronym{psf}{PSF}{Point Spread Function}
\newacronym{rf}{RF}{Radio Frequency}
\newacronym{rfi}{RFI}{Radio Frequency Interference}
\newacronym{roach}{ROACH-I}{Reconfigurable Open Architecture Computing Hardware}
\newacronym{ska}{SKA}{Square Kilometre Array}
\newacronym{snr}{SNR}{Signal to Noise Ratio}
\newacronym{udp}{UDP}{User Datagram Protocol}

\title[Direct-Imaging and FX Correlator for the BEST-2 Array]{Implementation of a Direct-Imaging and FX Correlator for the BEST-2 Array}
\author[G. Foster, J. Hickish, A. Magro, D. Price, and K. Zarb Adami]{G. Foster$^{1,3}$, J. Hickish$^{1}$, A. Magro$^{2}$, D. Price$^{1,4}$, K. Zarb Adami$^{1,2}$\\
$^{1}$University of Oxford, Sub-Department of Astrophysics, Denys Wilkinson Building, Keble Road, Oxford, OX1 3RH, United Kingdom\\
$^{2}$University of Malta, Department of Physics, Msida, Malta\\
$^{3}$Rhodes University, Department of Physics, Grahamstown 6140, South Africa\\
$^{4}$Harvard-Smithsonian Center for Astrophysics, 60 Garden Street, Cambridge, Massachusetts, 02138, USA}

\begin{document}

\date{\today}

\pagerange{\pageref{firstpage}--\pageref{lastpage}} \pubyear{2014}

\maketitle

\begin{abstract}
A new digital backend has been developed for the BEST-2 array at Radiotelescopi di Medicina, INAF-IRA, Italy which allows concurrent operation of an FX correlator, and a direct-imaging correlator and beamformer.
This backend serves as a platform for testing some of the spatial Fourier transform concepts which have been proposed for use in computing correlations on regularly gridded arrays.
While spatial Fourier transform-based beamformers have been implemented previously, this is to our knowledge, the first time a direct-imaging correlator has been deployed on a radio astronomy array.
Concurrent observations with the FX and direct-imaging correlator allows for direct comparison between the two architectures.
Additionally, we show the potential of the direct-imaging correlator for time-domain astronomy, by passing a subset of beams though a pulsar and transient detection pipeline.
These results provide a timely verification for spatial Fourier transform-based instruments that are currently in commissioning.
These instruments aim to detect highly-redshifted hydrogen from the Epoch of Reionization and/or to perform widefield surveys for time-domain studies of the radio sky.
We experimentally show the direct-imaging correlator architecture to be a viable solution for correlation and beamforming.
\end{abstract}

\begin{keywords}
instrumentation: interferometers ---
instrumentation: miscellaneous ---
techniques: interferometric ---
radio continuum: general
\end{keywords}

\glsresetall
\label{firstpage}

\section{Introduction}

We present a new digital backend designed for the \gls{best2} array at the Radiotelescopi di Medicina, INAF-IRA, Italy.
This backend is implemented on \gls{fpga}-based hardware from the \gls{casper}\footnote{\url{https://casper.berkeley.edu}} \cite{casper_ref}.
It consists of an FX correlator and a spatial \gls{fft} processor which can be used as a direct-imaging correlator or as a beamformer, supplying high time-resolution data to a pulsar and transient detection pipeline.

Within radio astronomy, sensitivity, field of view, and resolution requirements drive an upwards trend in the size of correlators.
In recent years many new arrays (for example: PAPER\footnote{\url{http://eor.berkeley.edu}}, MWA\footnote{\url{http://www.mwatelescope.org}}, LEDA\footnote{\url{http://www.21cmcosmology.org}}, LOFAR\footnote{\url{http://www.lofar.org}}) have been deployed with the goal of detecting highly redshifted hydrogen in the early universe.
These low frequency arrays are made up of $O(100)$ cross-correlated antenna elements (or beamformed ``stations''); their sensitivity is limited by total collecting area of these elements.
Increasing the number of antennas per array without a reduction in field of view requires a larger correlator, the cost of which scales as $O(N^2)$, where $N$ is the number of stations/elements.
Next-generation telescopes, such as the HERA-II array\footnote{\url{http://www.reionization.org}}, and the  \gls{ska}\footnote{\url{http://www.skatelescope.org}} will require order of magnitude larger $N$-correlators, and correspondingly the computational requirements of their correlators are significantly higher.

Where arrays have stations which are regularly distributed, direct-imaging correlators can take advantage of array redundancy to reduce correlation cost scaling to $O(N \log N)$.
This is achieved by directly producing images of the sky using multi-dimensional spatial \glspl{fft}.
By zero-padding the spatial \gls{fft} inputs, a correlation matrix of all unique $uv$ samples can be recovered after time integration by applying an inverse Fourier transform to this image.
Though the idea of direct image-formation by spatial \gls{fft} is not new in astronomy (see, for example \cite{2dfft}) retention of all $uv$ data by prior zero-padding is a relatively recent proposal \cite{fftt, omniscope}.
The \cite{omniscope} method has been extended, at an increase in computational cost, to accommodate non-regular arrays in \cite{moff} and \cite{bunton}.
To our knowledge, this BEST-2 backend represents the first deployment of a direct-imaging telescope which implements the design proposed by \cite{fftt}.

For \gls{eor} power spectrum studies, there are advantages in carefully designed maximally-redundant arrays, as is discussed in \cite{paper_array}.
PAPER, CHIME\footnote{\url{http://chime.phas.ubc.ca}}, and MiTEoR\footnote{\cite{mit_eor}} are all regularly-gridded arrays, and regularly-gridded layouts are being considered for the HERA-II array.
Nevertheless, it must first be verified that direct-imaging correlators can deliver calibrated science-quality data products.

A challenge of the direct-imaging correlator architecture is that calibration solutions for individual antennas must be applied during the observation; they cannot be applied post-correlation as the spatial \gls{fft} effectively averages all redundant baselines together.
By developing a direct-imaging correlator, we have explored the calibration challenge in a practical way.
We show that with a stable system the direct-imaging correlator, once calibrated, can produce viable images, visibilities, and beams to be used for scientific observation. 

The rest of this section provides a review of how the spatial Fourier transform can be used as the basis for the direct-imaging correlator.
In addition, we include an overview of the BEST-2 array, which serves as a suitable direct-imaging frontend.
Section \ref{section:firmware} is a detailed overview of the \gls{fpga}-based system on which the BEST-2 backend is implemented.
In Section \ref{section:obs}, we present the results of initial observations with the backend, including images from both the direct-imaging and FX correlator systems, and a preliminary pulsar observation.

\subsection{Correlation using a spatial Fourier transform}
\label{sec:corr_using_fft}

An FX correlator computes the Fourier transform of time-domain antenna signals (the \emph{F} operation) and then forms $uv$ samples by per-channel, per-sample cross-multiplication of all antenna signal pairs (the \emph{X} operation). For an antenna array with $N$ elements, there are $O(N^2)$ such combinations of antenna pairs, which gives rise to an $O(N^2)$ computational cost of FX correlators.

The formation of $uv$ samples by cross-multiplication is equivalent to a spatial autocorrelation of channelized antenna signals. 
This can be implemented by performing a per-channel, per-sample spatial Fourier transform on the channelized signals from the antennas in an array, then taking the element-wise power of the result, and finally inverse Fourier transforming the result (after averaging over time) to recover the cross-power spectrum. 
The direct-imaging correlator is a correlator that performs precisely these steps, and is so named because, prior to the final inverse Fourier transform, the intermediate data product is an image of the sky. 
In this sense, a direct-imaging correlator is not really a correlator at all, but rather a beamformer capable of forming enough beams simultaneously to obtain a complete image of the sky.

In general, the largest number of independent $uv$ measurements which can be made by an array of $N$ elements is of order $N^2$. The direct imaging correlator provides an opportunity to perform these measurements by forming an equivalent number of beams on the sky rather than computing correlations directly.

When $N$ receiving elements in an antenna array are placed on a regularly spaced grid, only $O(N)$ independent $uv$ measurements are obtained, since many of the two-element interferometers in the array are redundant. It is thus possible to encapsulate all of these measurements in $O(N)$ beams. These could be formed by simple discrete Fourier transform and power-detection, which has a computational cost of $O(N)$ operations per beam. However, a well-known method for producing a complete set of orthogonal beams on the sky from regularly spaced antennas is by spatial \gls{fft} beamforming \cite{fastbeamforming}.
Such a beamforming implementation is able to generate $N$ beams on the sky, at a computational cost of $O(N\log{N})$.
For large-$N$ arrays, this can be a significant computational saving compared to imaging by FX correlation which requires $O(N^2)$ operations to compute all $O(N^2)$ pairwise antenna correlations.

Essentially the only difference between the imaging methods outlined by \cite{fftt} and the FFT beamforming of \cite{fastbeamforming} is that the former requires zero padding to be applied to the matrix of antenna signals before the spatial \gls{fft} is performed. In a physical sense, this ensures that enough (real-valued) beam powers are measured to encode all the (complex-valued) $uv$-plane information which would be measured by a correlator.
The zero-padding required results in the generation of $~2^{m}N$ beams on the sky where $m$ is the number of dimensions in the antenna array. In the \gls{best2} case $m=2$.

It is worth reiterating what a $uv$ measurement represents in terms of the direct-imaging correlator.
Instead of making individual measurements of redundant two-element interferometers in an array (as in an FX correlator) the direct-imaging correlator computes an ``effective" visibility which is the average of the visibilities in a redundant set of interferometers.
But, for the average visibility to effectively represent the sky $uv$ mode being measured, each redundant baseline measurement must be identical.
Thus, calibration to the complex gains must be applied before the direct-imaging operation.
Further calibration can be applied after imaging, but comes with the penalty of reduced signal to noise.
We will consider what implications this has on observations in a future work comparing the FX and direct-imaging correlators.

\subsection{BEST-2 Array}

\begin{figure*}
    \centering
    \subfigure[
    ]{
        \includegraphics[width=.46\textwidth]{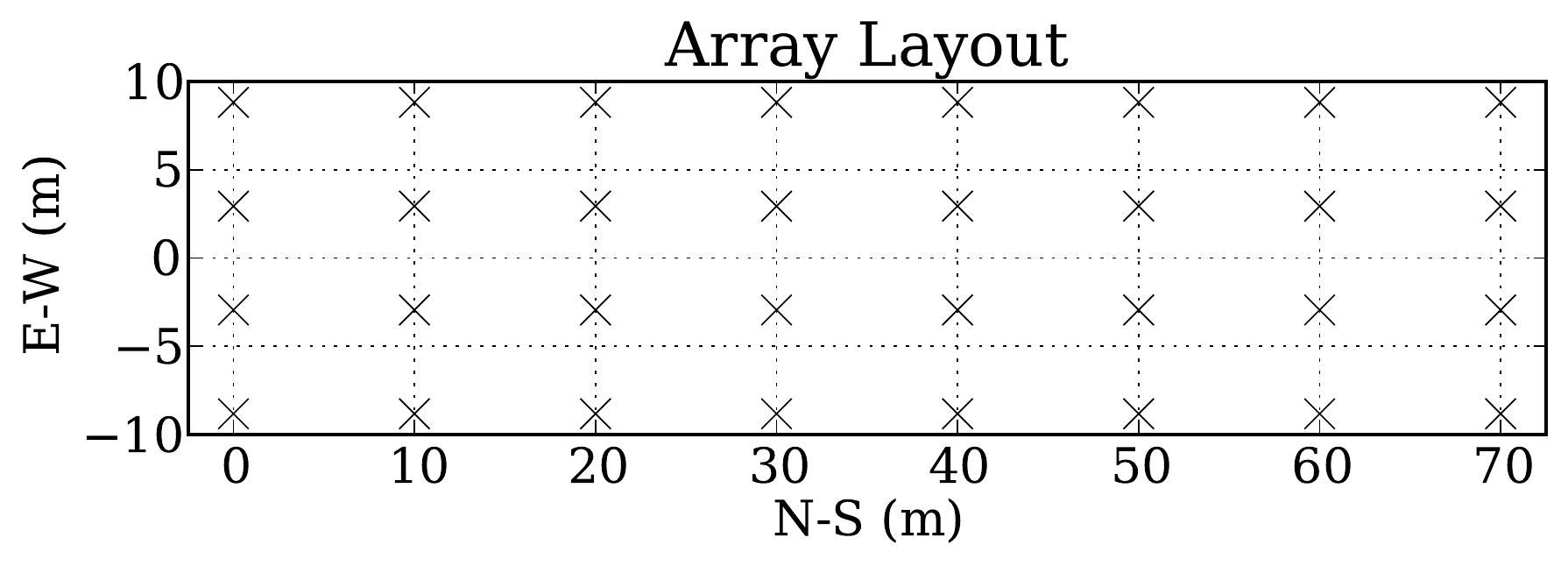}
        \label{fig:ant_layout}
    }
    \hspace{3mm}
    \subfigure[
    ]{
        \includegraphics[width=.46\textwidth]{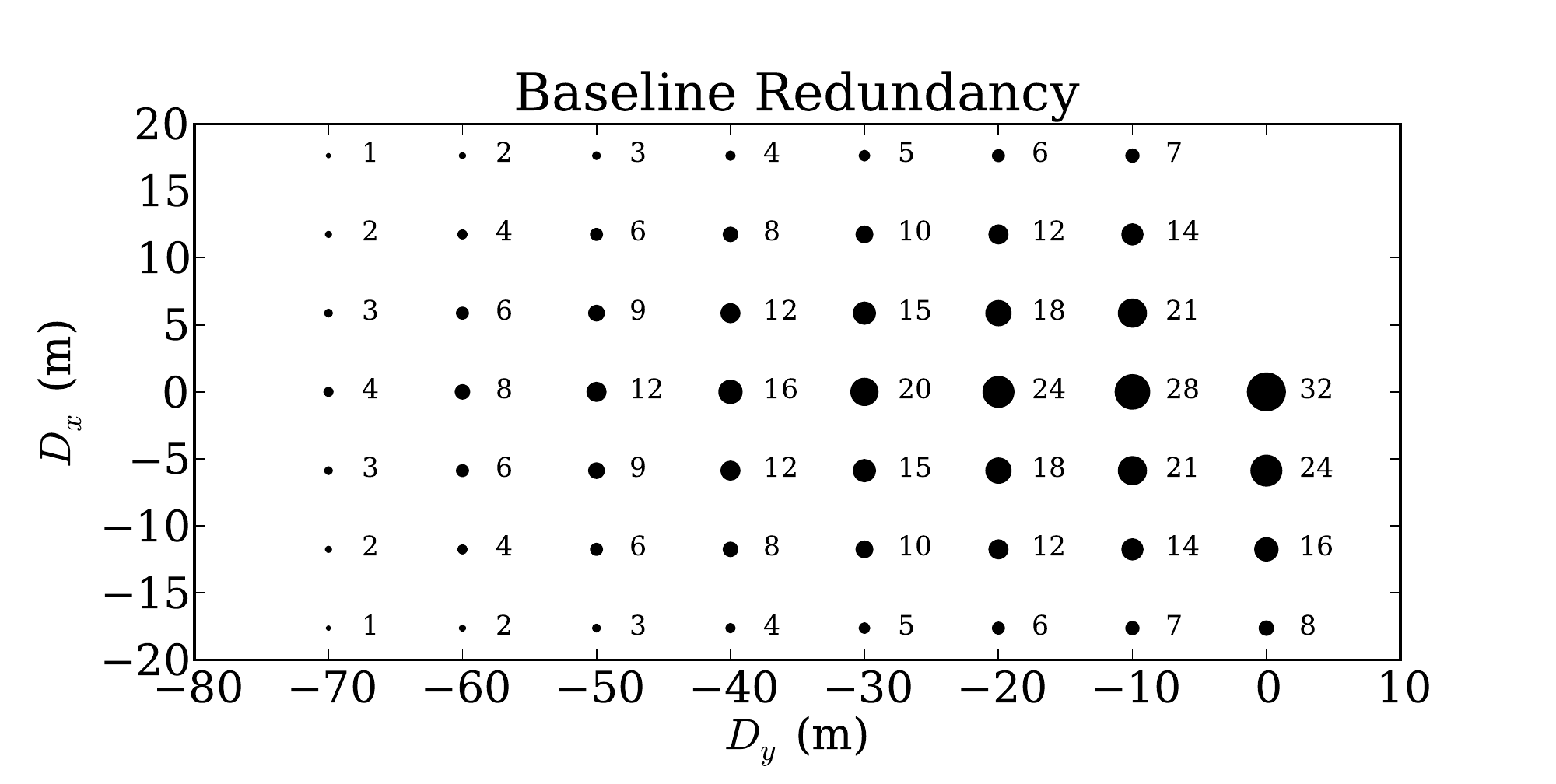}
        \label{fig:red_bl}
    }
    \caption{BEST-2 array configuration.
    Figure \subref{fig:ant_layout} is the array layout of the eight cylindrical antennas each with four effective receivers of 16 beamformed dipoles. The array is on a \emph{4-by-8} grid.
    Figure \subref{fig:red_bl} shows the array baseline redundancy, circle radius shows the relative baseline redundancy, values indicate number of redundant baselines for a given length.
    There are 53 unique baselines.
    This is also the effective \emph{uv} coverage for a source at the local meridian.
    }\label{fig:best2_layout}
\end{figure*}

The \gls{best2} array consists of 8 East-West oriented cylindrical concentrators, each with 64 dipole receivers critically sampling a focal line at 408 MHz.
Signals from these 64 dipoles are analogue beamformed in groups of 16, resulting in 4 receiver channels per cylinder, and a total of 32 effective receiving elements laid out on a $4 \times 8$ grid, depicted in Figure \ref{fig:ant_layout}.
The cylinders, which lie on an east-west axis, can be rotated to point at a constant declination within the array pointing limits.
A synopsis of the \gls{best2} array specifications is presented in Table \ref{tbl:best2}.
\gls{best2} was developed as a reliable, low cost analogue front end to be used in \gls{ska} development, with a core design requirement of simplicity in interfacing with different digital backends \cite{best2}.
Extensive documentation of the development of the \gls{best2} analogue chain can be found in several earlier publications (\cite{best2-lna}; \cite{best2-rec}).

\begin{table}
      \centering
      \begin{tabular}{| l | l | l |}
      \multicolumn{2}{|c}{BEST-2 Array Specifications}\\
      \hline
      Total Number of receivers 		&         32 \\
      Total Collecting Area 		&     $1411.2 \: m^2$ \\
      $A_{eff}/T_{sys}$ 		&      $11.65 \: m^2/K$ \\
      Longest Baseline (E-W) 	&      17.04 m \\
      Longest Baseline (N-S) 	&      70.00 m \\
      Central Frequency 		&        408 MHz \\
      Analogue Bandwidth 		&         16 MHz \\
      Primary Beam Area 	 		&      58.6 $\textrm{deg}^2$ \\
      \hspace{5 mm} North-South 			&        6.86$^{\circ}$ \\
      \hspace{5 mm} East-West 		&        8.77$^{\circ}$ \\
      Zenith-Pointing PSF Area 				&        $\frac{1}{3} \: \textrm{deg}^2$ \\
      \hspace{5 mm} North-South 			&       16' \\
      \hspace{5 mm} East-West 		&       75' \\
      \end{tabular}
      \caption{
      The top level specifications of the BEST-2 Array, a subset of the Northern Cross, located near Medicina, Italy.
      The array is in a regularly gridded $4 \times 8$ layout.}
      \label{tbl:best2}
\end{table}

The gridded layout of the array is a key requirement of the direct-imaging correlator architecture.
This grid structure leads to a highly redundant array, Figure \ref{fig:red_bl}, where there are $\sim 2N$ unique baselines.
A direct-imaging correlator architecture produces all the unique baselines where any `effective baseline' is the average of all the redundant baselines for that baseline length and projection.
The gridded layout also produces regular \emph{uv} coverage, Figure \ref{fig:red_bl}, of a field which leads to strong sidelobe in the \gls{psf} which can be seen in the images.

\section{FPGA Instrumentation}
\label{section:firmware}

The new \gls{fpga}-based instrumentation we have developed has replaced the previous generation FX correlator system which used \gls{casper} iBOB and BEE2 boards, \cite{best2-casper}.
The new instruments have been implemented on \gls{fpga}-based ROACH boards also developed by \gls{casper}.
The ROACH platform is based on a XILINX Virtex 5 SX95T 
\gls{fpga} with interfaces to DRAM and QDR memory, high speed CX-4 connectors, and a generic Z-DOK interface for connecting ADCs and various other daughter boards.
\gls{dsp} components of the firmware were developed using MATLAB Simulink, the \gls{casper} \gls{dsp} libraries, and custom \gls{hdl} code.
Firmware models, bitstreams, and control software are available from our project repository\footnote{\url{https://github.com/griffinfoster/medicina}}.

The instrument is made up of three interconnected ROACH boards.
These will be referred to as the F-Engine, S-Engine, and X-Engine for their respective functions.
Signal digitization and frequency channelization is performed on the F-Engine.
Duplicate streams are sent to the X-Engine for cross-correlation, and to the S-Engine where a spatial Fourier transform is performed.
Table \ref{tbl:dig_instrument} lists the digital specifications of the implemented \gls{fpga} hardware.
Block diagrams of the firmware for each board are included in a supplemental online appendix.

\begin{table}
      \centering
      \begin{tabular}{| l | l | l |}
      \multicolumn{2}{|c}{Digital Hardware Specifications}\\
      \hline
      Digital Bandwidth  & 20 MHz \\
      ADC Sampling      & 12-bit \\
      PFB-FIR           & 4-tap Hann Window \\
      PFB-FFT           & 2048 radix-2 \\
      Freq. Channel Width& $19.53$ kHz \\
      Quantization      & 4-bit complex \\
      X-Engine          & \\
      \hspace{5mm} Auto Correlations    & 32\\
      \hspace{5mm} Cross Correlations   & 496\\
      \hspace{5mm} min $\tau_{int}$     & $6.55$ ms\\
      S-Engine          & \\
      \hspace{5mm} 2-D FFT          & $8 \times 16$ \\
      \hspace{5mm} Selectable Beams     & 128 \\
      \hspace{5mm} Unique Baselines     & 53 \\
      \hspace{5mm} min $\tau_{int}$     & 1 s\\
      \end{tabular}
      \caption{Specification of the digital instrument made of the F-Engine, S-Engine, and X-Engine boards which implement the direct-imaging and FX correlator architectures.
      }
      \label{tbl:dig_instrument}
\end{table}

\subsection{F-Engine}

Both correlators use the same digitization and channelization frontend.
The F-Engine processing node is responsible for digitization, channelization, and transmission of quantized antenna signals to downstream processing boards. 
Duplicate streams are sent from the F-Engine to the S-Engine and X-Engine so that the correlators may be used concurrently.
This allows for a streamlined process for calibrating the direct-imaging correlator data and simultaneous observation with both correlators, and reduces the amount of hardware.

\subsubsection{Digitization}

In the analogue chain the \gls{rf} signal, centered at 408~MHz, is mixed down with an 378~MHz \gls{lo} to an \gls{if} of 30 MHz.
Prior to digitization the last amplifier stage of the analogue chain has per-signal channel adjustable gain useful for setting levels for optimum \gls{adc} quantization.
Signal digitization is performed using the CASPER 64ADCx64-12\footnote{\url{https://casper.berkeley.edu/wiki/64ADCx64-12}} \gls{adc} board which makes use of the Texas Instruments ADS5272 8-channel, 12-bit \gls{adc} chip.
In the case of the 64ADCx64-12, eight ADS5272 \gls{adc} chips are used to digitize 64 analogue inputs at up to 50~Msps.
The 64ADCx64-12 board interfaces with the ROACH over 64 \gls{lvds} pairs via Z-DOK.
In our design 32 signal streams are digitized at 40 Msps for 20~MHz of digital bandwidth which covers the 16~MHz analogue band.

\subsubsection{Channelization}

In order to facilitate the correlation and imaging processing described in Section \ref{sec:corr_using_fft}, antenna signals must be split into multiple frequency channels. These channels should have narrow enough bandwidth that they can be treated as monochromatic. This leads to the requirement that the relative arrival time of radiation at different antennas in the array should be much shorter than the inverse bandwidth of each channel. In the most stringent case of radiation arriving from the horizon we require
\begin{equation}
    \frac{1}{B} = \alpha\frac{D}{c},\qquad\qquad \alpha\gg1\,,
\end{equation}
where $B$ is the bandwidth of a frequency channel, $D$ is the largest dimension of the array, and $c$ is the speed of light. For the BEST-2 array, $D=70$ m which with $k=10$ sets a minimum channel bandwidth of approximately 400~kHz.

The strong \gls{rfi} environment at the observatory sets a more restrictive demand for narrow channel widths and high out of band rejection that is achieved by use of a \gls{pfb}.
Our digitized band is channelized using a critically-sampled 4-tap Hann filter, 2048-point \gls{pfb} (using 18-bit coefficients) to produce 1024 complex samples per real antenna stream covering the range 398--418~MHz.
Individual channel resolution is $19.53$ kHz.
Since the \gls{roach} board is clocked at 160 MHz; four times the \gls{adc} sample rate, such that four signals are time-division multiplexed onto a single signal path.

\subsubsection{Equalization and Quantization}

After channelization the samples are requantized to 4-bit complex values.
This limits the F-Engine output data rate to 5.12 Gbps.
See Section 8.3 of \cite{Thompson:2004fk} for details on the effect of quantization noise in a 4-bit system.

The characteristics of a telescope analogue front-end mean that a compensation of the instrumental bandpass must be applied to data to ensure optimal quantization of all subbands. 
In the BEST-2 backend, this equalization is achieved by pre-multiplying the 18-bit data streams by a set of 18-bit coefficients, with 512 independent coefficients available to represent the bandpass of each analogue channel. 
During observations, an automated script samples the pre-quantized signals and statistically determines the correct equalization coefficients to achieve optimal 4-bit quantization.
This script can be run at a user selected cadence during observations.

The per-channel equalizers are also used to apply pre-derived amplitude and phase corrections to the signals before the 2-D \gls{fft}.
These values can by dynamically set during observations to account for changes in the system calibration.
Duplicate streams, though with different ordering, are sent over XAUI at a rate of 5.12 Gbps to the X-Engine and S-Engine nodes.

\subsection{X-Engine}

An FX correlator architecture is a standard design for large bandwidth, many antenna arrays.
An overview of the architecture is presented in \cite{romney_white_book}.
The core component to the \emph{X} stage of the FX correlator is the \gls{cmac} of all pairs of independent signals for each frequency channel.
A pipelined X-Engine, based on the general \gls{casper} block \cite{fxcorrelator}, is used for optimal multiplier efficiency.

The correlation is performed on a per channel basis, thus the channelized band can be split up into portions and processed in parallel across multiple X-Engines.
This allows a larger bandwidth to be processed at the cost of increased logic and multiplier resource utilization.
In the case of the \gls{best2} system two pipelined X-Engines, each processing 512 channels, are implemented in the firmware.

A two-stage accumulator is used to reduce the output data rate.
The first stage is fixed to give a minimum integration time to 6.55 ms; this assures that the output bandwidth to each X-Engine is approximately equal to the input bandwidth.
The second accumulator, implemented using QDR, is software-controlled with integration lengths up to many minutes.
A completed correlation matrix is sent to a receive computer over a \gls{10gbe} connection.
Correlation matrices are packetized as \gls{udp} streams using the SPEAD protocol\footnote{https://github.com/ska-sa/PySPEAD} .

\subsection{S-Engine}

In the BEST-2 backend described here, the requirements on the direct-imaging correlator are multi-fold.
The system should be capable of generating images on order second timescales, by the method described by \cite{fftt}.
Further, the system should be capable of passing formed beams at full bandwidth, i.e. without any accumulation, to downstream time-domain processing systems such as a real-time pulsar dedispersion engine, \cite{dedispersion}. 

Using a single frequency channel from each antenna element and the geometric array layout a two-dimensional \gls{fft} is performed to produce an image.
The two-dimensional \emph{8-by-16} \gls{fft} is performed using an 8-point parallel input \gls{fft} followed by a corner-turn and 16-point parallel input \gls{fft}.
Zero padding is added to the signals from the $4 \times 8$ antenna \gls{best2} array before input into each stage of the transform.
This yields an output of 128 complex-valued beams, from which the power is calculated and accumulated.
Each channel is cycled over to produce an image for each frequency band.
The four-fold increase is the number of outputs from the spatial transform by zero padding introduces a number of redundant calculations.
The spatial transform produces 128 outputs but, as shown in Figure \ref{fig:red_bl}, there are only 53 unique baselines in a  $4 \times 8$ grid.

Accumulation takes place in two-stages.
A fixed 128-sample vector accumulator reduces the output of the \gls{fft} by summing 128 consecutive time samples of each beam for a single frequency channel.
The complete set of 1024 channels and 128 beam powers can then be multiplexed into a single software-controllable vector accumulator in off-chip memory.
Accumulations are packetized by the ROACH's PowerPC co-processor and sent to a data recorder over 1 GbE link using the same protocol as the FX correlator. 

\subsubsection{Beamformer}

A total of 128 beams are generated with the spatial Fourier transform.
Before calculating beam powers, 8 arbitrary beams can be selected and output over \gls{10gbe} at 16-bit precision for time-domain processing.
Packets are constructed from multiple time and frequency samples from individual beams such that processing can be parellelized by beam across multiple machines.
For a data word size $W$ in bytes, a packet containing $T$ samples and $C$ channels has payload size $P=W \times T \times C$ bytes.
In the \gls{best2} backend, $W = 4$, and $T = 128$ is chosen to conveniently match the accumulation length of the imager. 
$C = 8$ is chosen to achieve packets with payloads of 4096 kB. 
These are generated with a simple data transpose in QDR memory space, and headers are inserted before transmission over \gls{10gbe}.
A pulsar and transient searching system for this output has been developed on \glspl{gpu} by \cite{multibeamgpu}.

\begin{figure}
    \includegraphics[width=.46\textwidth]{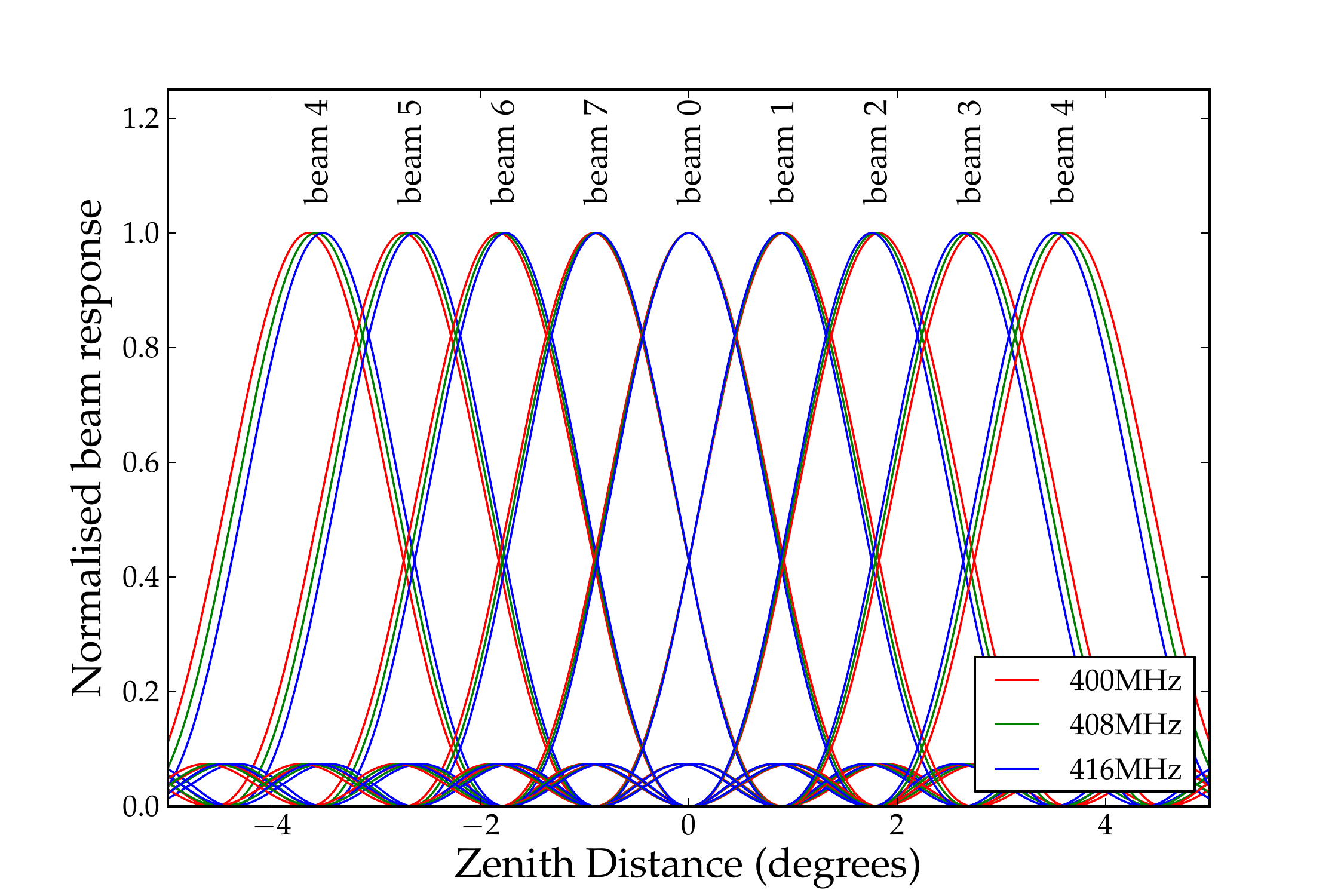}
    \caption{Theoretical beam responses (with aliases outside the primary beam removed) for an East-West strip of eight beams at 400, 408, and 416 MHz. Beam 4 is plotted twice, as its aliases give rise to equal beam response both East and West of the zenith.}
    \label{fig:beam_centres}
\end{figure}

Whilst the electric-field beams generated by spatial Fourier transform are suited to ``drift-scan" observations, software has been provided to switch beams to ``psuedo-track" up to 8 simultaneous targets as they transit the primary beam.
In this mode of operation, software will dynamically configure the S-Engine node outputs to transmit the nearest beam to a target source. 
This tracking method has the drawback that sources appearing between beam centers can not be observed with the full sensitivity of the array. 
Figure \ref{fig:beam_centres} shows the response of the 7 synthesized beams covering a right-ascension range of $4^{\circ}$. 
Since neighbouring beams overlap at $\sim 81 \%$ of peak power, signal-to-noise is reduced when tracking objects lie between pointing directions. 
Additionally, the beam centers vary with frequency. 
Since the fractional bandwidth of the \gls{best2} analogue frontend is relatively small, this effect is limited, especially around the pointing centre of the telescope. 
The beams around the pointing-centre are also preferred as they suffer least from spatial aliasing and primary beam gain attenuation.

\section{Initial Observations}
\label{section:obs}

\begin{figure}
    \centering
        \includegraphics[width=.44\textwidth]{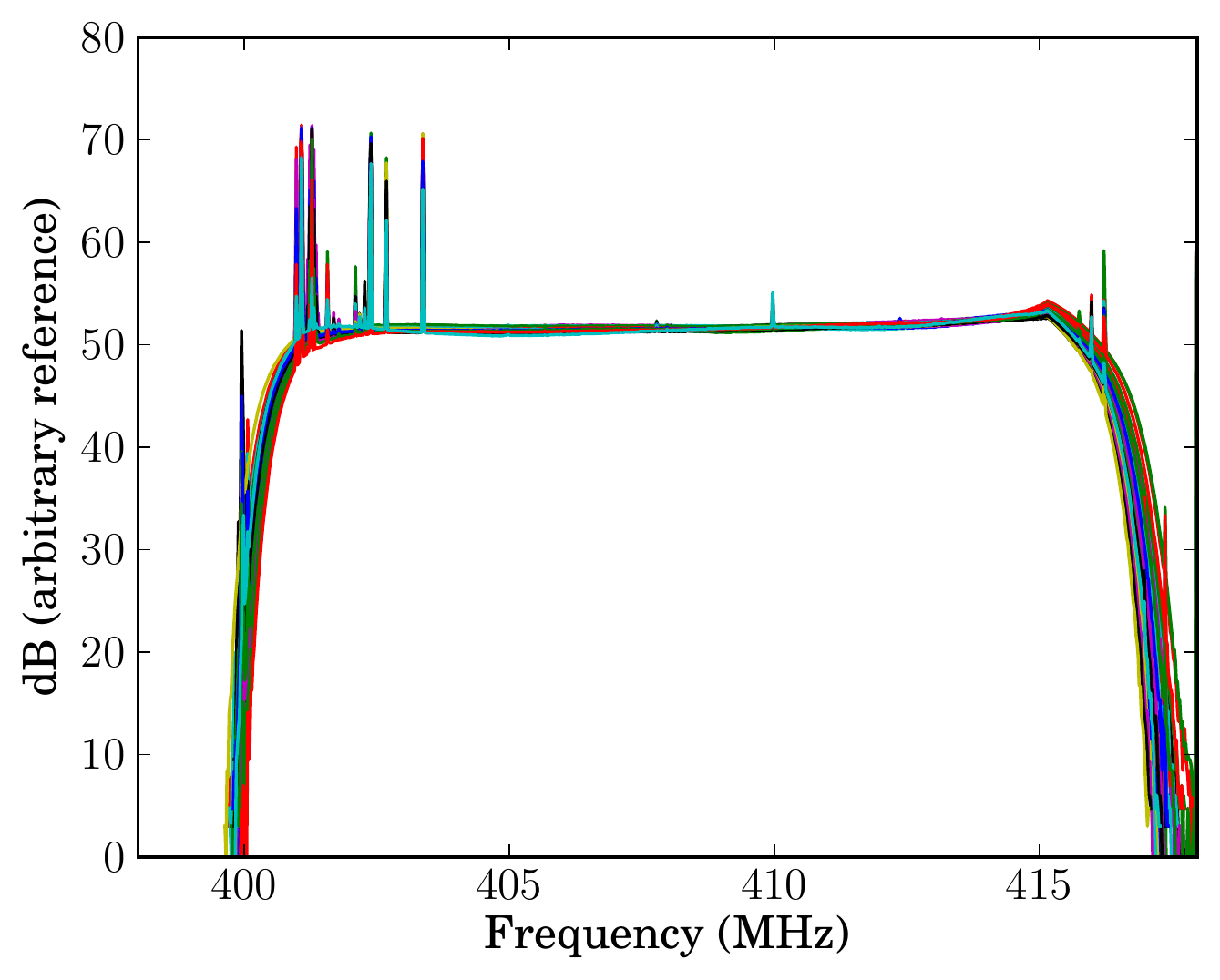}
        \label{fig:autos_post}
    \caption{Autocorrelation spectra of 32 BEST-2 receivers after application of bandpass gain corrections.}\label{fig:autos}
\end{figure}

Instrumentation was installed and tested in March 2012.
Initial tests verified the F-Engine firmware during observations.
Figure \ref{fig:autos} shows the autocorrelation spectra of the 32 receivers of the \gls{best2} system after (Figure \ref{fig:autos_post}) quantization equalization is applied to account for gain differences and band slope. 

After installation, the FX correlator was used for observations of various bright radio sources.
Since the Northern Cross is a transiting array, with primary beam pointing adjustable in declination, there is a limited period of time each day in which a source is in the primary beam.
Bright sources such as Cygnus A, Cassiopeia A, Taurus A, and a number of 3C sources were observed.
Additionally, observations of multiple constant declination 24-hour cycles were made.

In concurrency with the FX correlator observations, the direct-imaging correlator recorded data, initially in a ``raw" mode where no pre-calibration was applied and then in normal observational mode with complex gains solutions from the FX correlator applied in real-time.

\subsection{FX Imaging}

\begin{figure}
    \centering

    \subfigure[3C10]{
        \includegraphics[width=.23\textwidth]{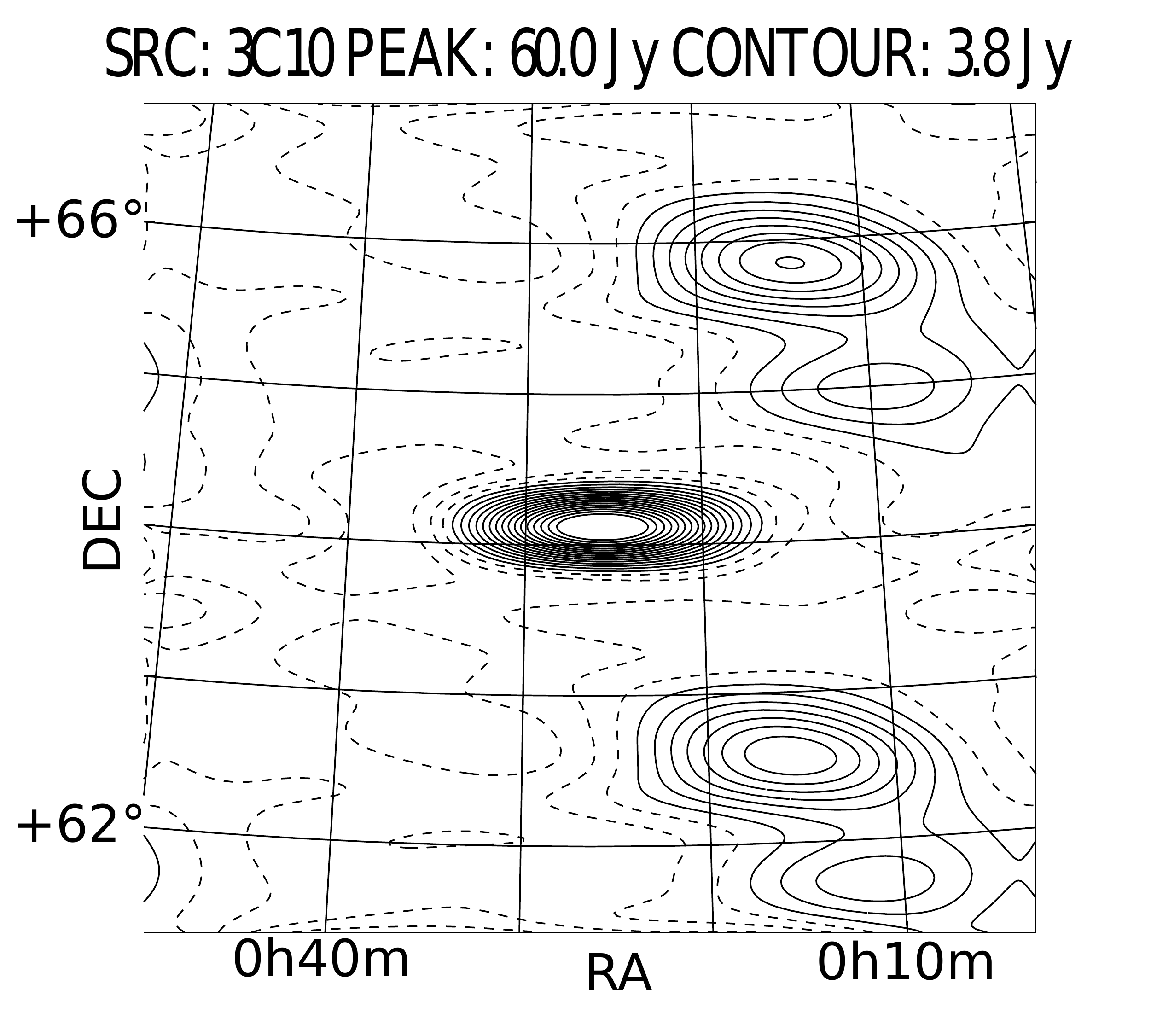}
        \label{fig:fx_3c10}
    }
    \hspace{-8mm}
    ~
    \subfigure[3C48]{
        \includegraphics[width=.23\textwidth]{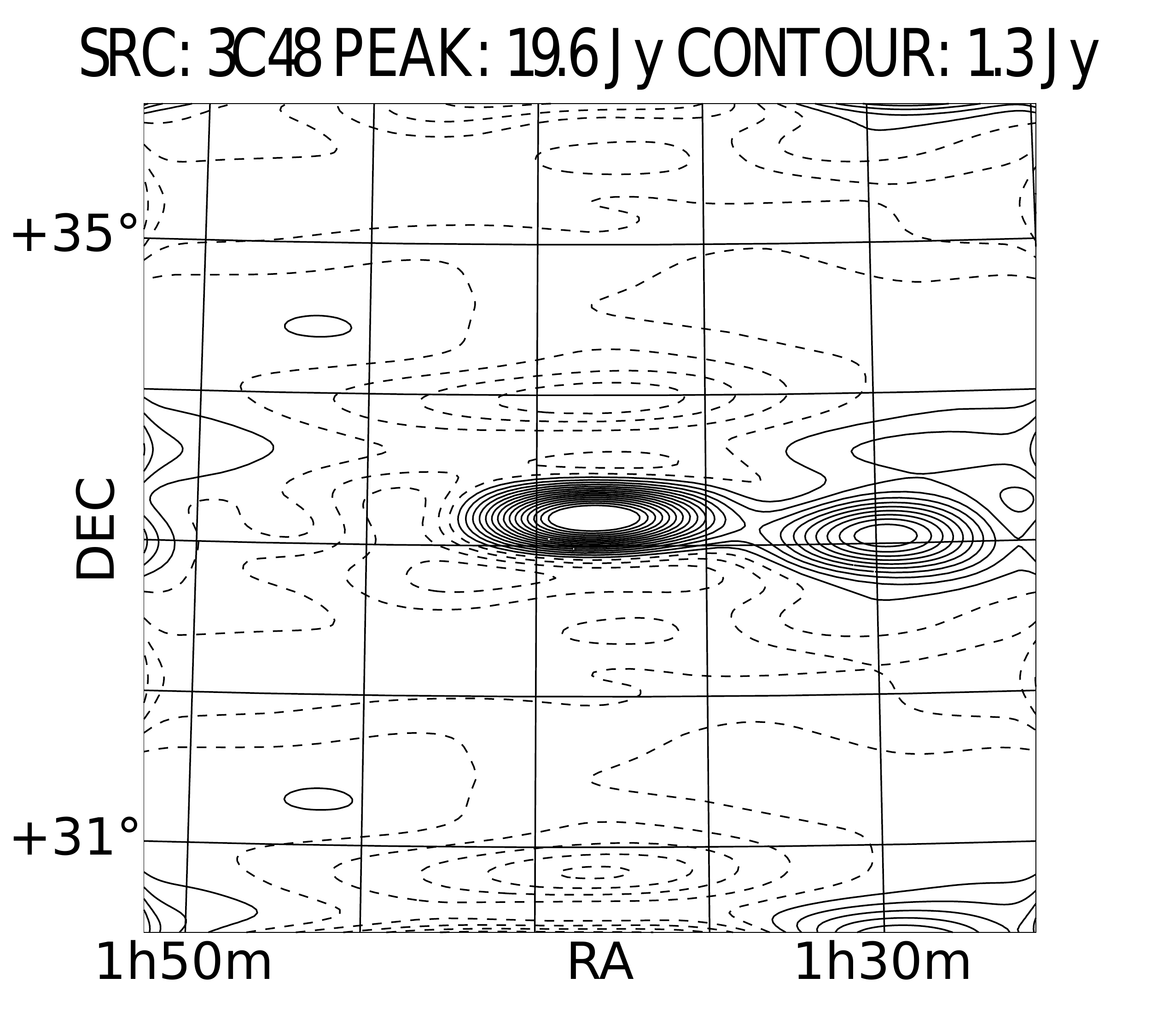}
        \label{fig:fx_3c48}
    }
    \vspace{-3mm}
    \subfigure[3C123]{
        \includegraphics[width=.23\textwidth]{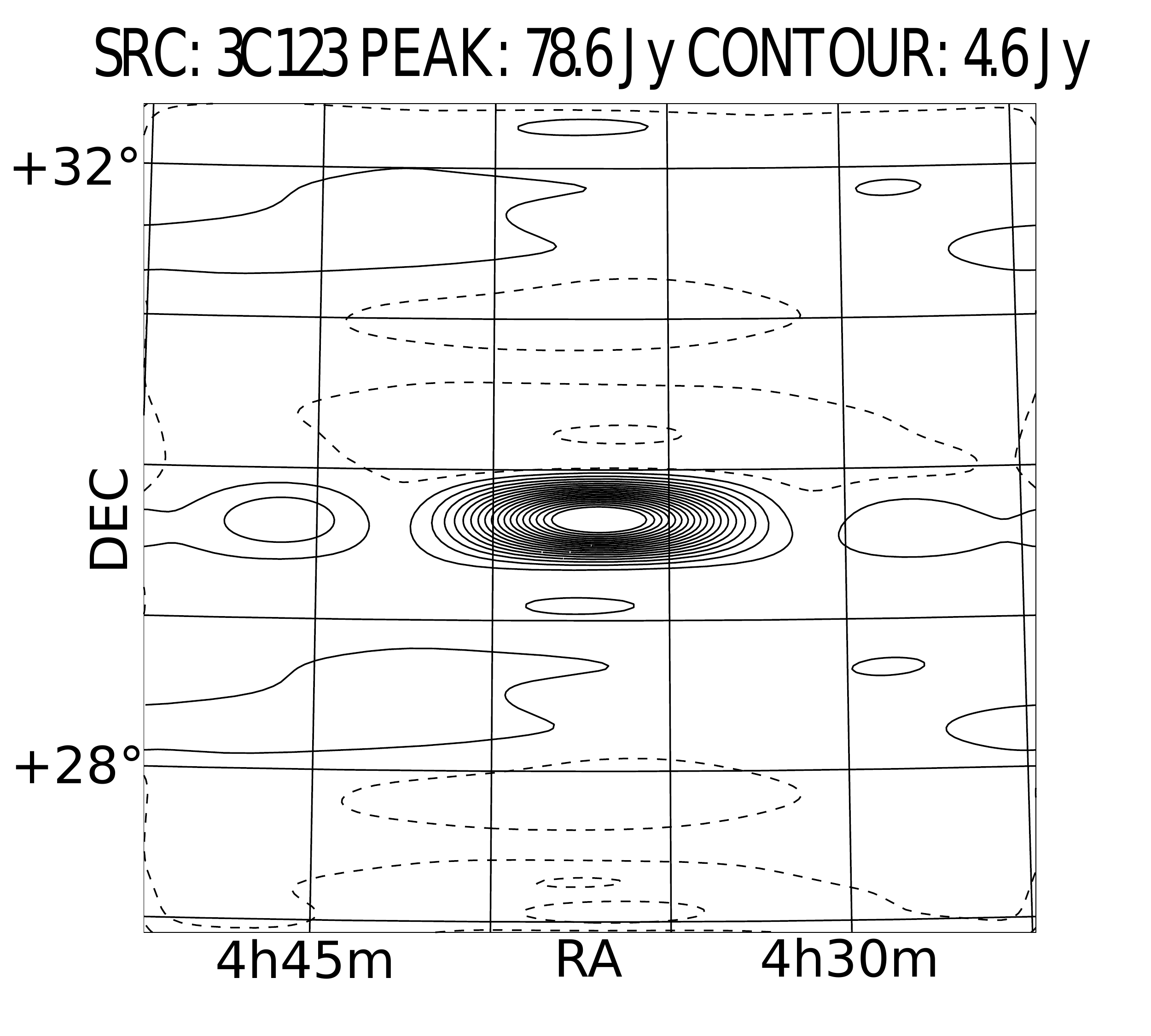}
        \label{fig:fx_3c123}
    }
    \hspace{-8mm}
    ~
    \subfigure[3C157]{
        \includegraphics[width=.23\textwidth]{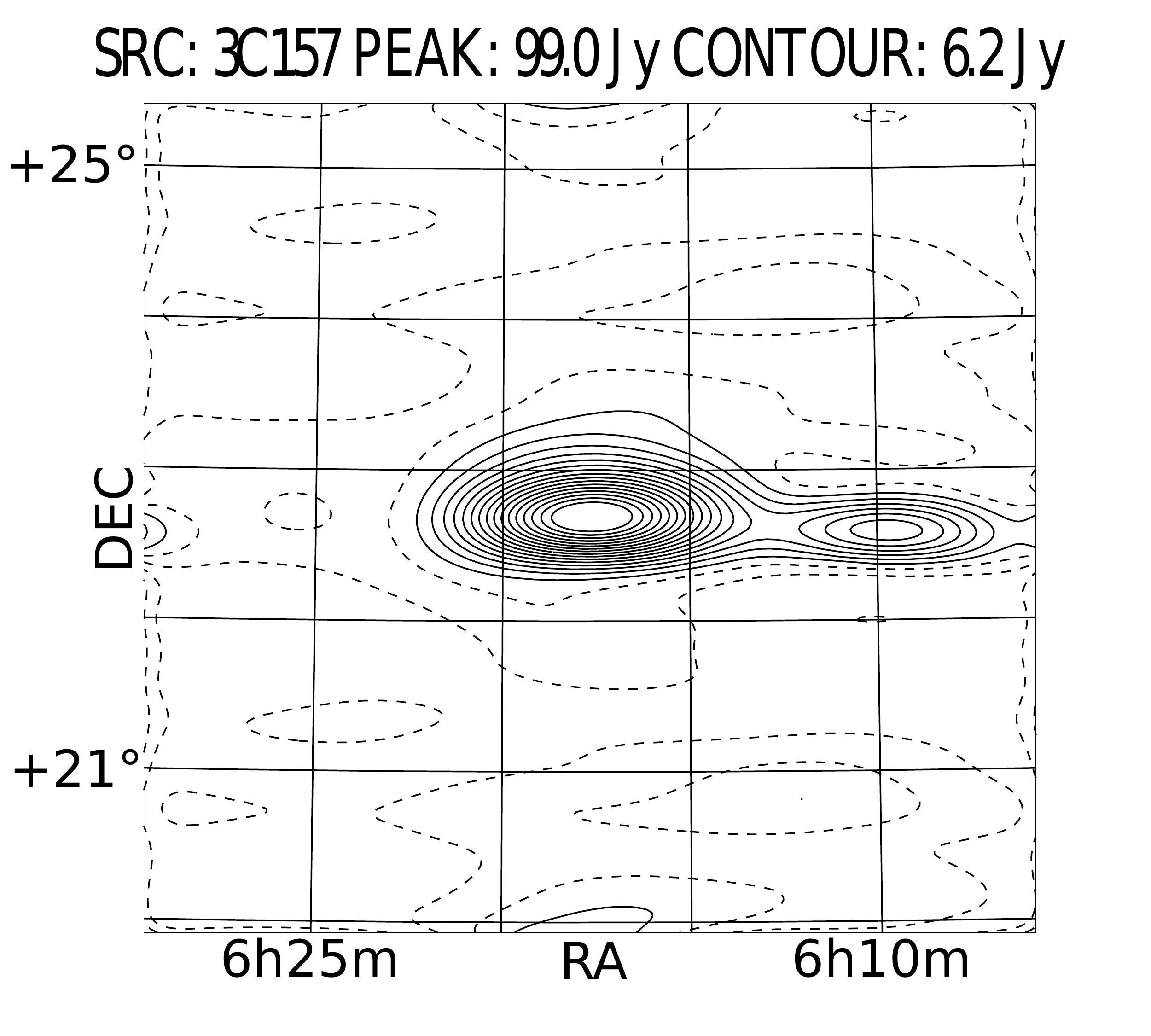}
        \label{fig:fx_3c157}
    }
    \caption{Calibrated `snapshot' images of various 3C fields produced using the \emph{FX} correlator, images have a noise level of $\sim 1.5$ Jy.
    }\label{fig:fx_images}
\end{figure}

A transiting array provides a unique challenge of gain calibration since a source's apparent gain will change as it transits the primary beam.
To account for this variation, a two stage gain calibration method is used.
Phase calibration and setting the flux scale is accomplished using Cassiopeia A and Cygnus A observations as point source sky models set to their \cite{Baars:1977uq} flux levels and known spectral indices.
Since these sources are very bright, only a few seconds at transit is needed produce a high \gls{snr} dataset to use for calibration.
Over this period the primary beam can be approximated as flat.
A time-independent complex gain is derived for an initial calibration.
After applying the gain corrections an observation will be set to a flux scale relative to the flux of the calibration source.
Each individual source is then calibrated in MeqTrees \cite{meqtrees} based on a local sky model taken from the 3C catalog.
A selection of images using this calibration method are shown in Figure \ref{fig:fx_images}.
An effect of the density of the antenna layout is a low field of view to \gls{psf} size ratio, $\sim 25$ in the north-south direction and $\sim 7$ in the east-west direction, thus images tend to contain at most a few spatially separated point sources.

\subsection{Direct Imaging}

The native output of the direct-imaging correlator is a 128 pixel image-domain dataset where each pixel is a unique beam within the field of view.
These 128 beams can be Fourier transformed to obtain samples in the \emph{uv} domain.
This produces summed visibilities corresponding to each of the 53 unique baselines, where each visibility is the average of all redundant baselines with a particular $uv$ spacing.
By transforming to the \emph{uv} domain we can compare the direct-imaging correlator data quality to that of our standard FX correlator.

\subsubsection{Complex Gain Calibration}

For our initial deployment the FX correlator was used to derive complex gain solutions which were then applied in to the direct-imaging correlator in realtime during subsequent observations.
Whilst the \gls{best2} direct imaging test-bed system utilizes the full correlator provided by the backend, it should be noted that an FX correlator which computes the full correlation matrix is not required for deriving gain calibrations.
In order to make the direct imaging architecture computationally beneficial, a smaller correlator, such as the calibration correlator included in LOFAR station firmware, or an offline correlator should be implemented.

In order to combine baseline data in a way that preserves signal-to-noise, a key requirement of the \gls{best2} direct-imaging system is the prior knowledge of gain corruptions in the receiving chains of each of the 32 receiving elements.
Given the regular layout of antennas in this array, redundant baseline calibration strategies such as those described by \cite{Noorishad2012} are applicable.
However, we opt for a general sky-model based algorithm which assumes prior knowledge of the sky observed by the telescope, but does not impose the condition that redundant baselines behave identically.
This strategy is chosen for a number of reasons.
The strong sidelobes of \gls{best2} primary beams, which are not identical even amongst redundant baselines, make calibration challenging when there are bright sources away from the phase-center of the array.
In such a situation, the assumption of redundant baselines breaks down and redundant-baseline calibration methods are likely to prove ineffective.
Even without requiring baseline redundancy, varying sidelobe gains across the array makes for challenging calibration as gain corruptions are strongly direction dependent.
To avoid this pitfall, the simplest calibration method is to wait for a transit of a bright source.
In this case, the model sky is trivially a point source at the array phase center from which gain solutions can be very easily extracted.

We use the column ratio gain estimation method, described by \cite{gaindecomp}.
Chosen for simplicity and ease of implementation, the algorithm uses ratios of antenna cross-correlations to estimate the diagonal (auto-correlation) elements of a measured visibility matrix in the absence of a system noise contribution.
This allows creation of a modified correlation matrix and can subsequently be used, under the assumption of a model sky consisting of a point source at phase center, to obtain per-antenna gain solutions.

\begin{figure}
    \centering

    \subfigure[Uncalibrated]{
        \includegraphics[width=.22\textwidth]{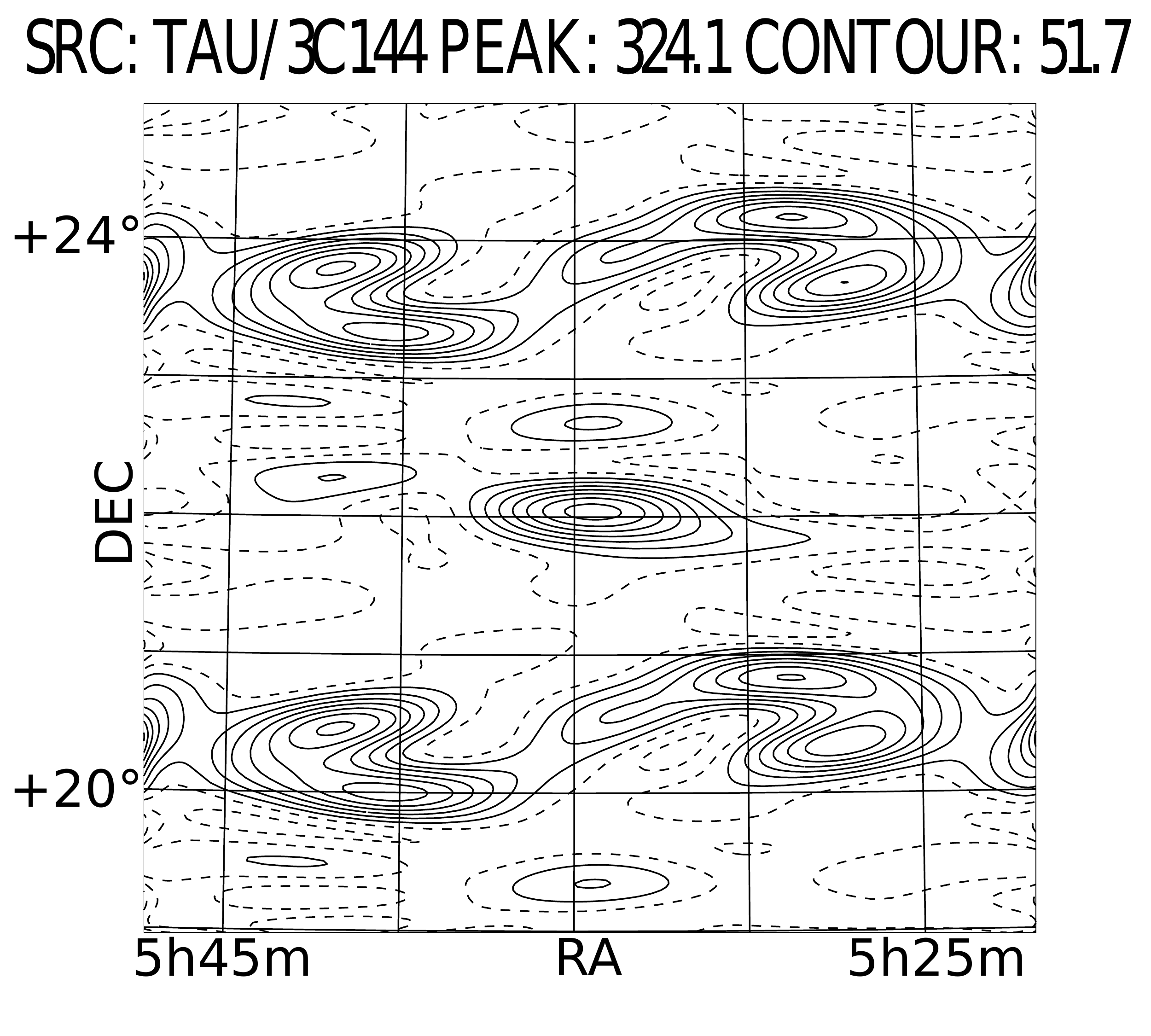}
        \label{fig:fs_uncal}
    }
    \hspace{-4mm}
    ~
    \subfigure[Calibrated]{
        \includegraphics[width=.22\textwidth]{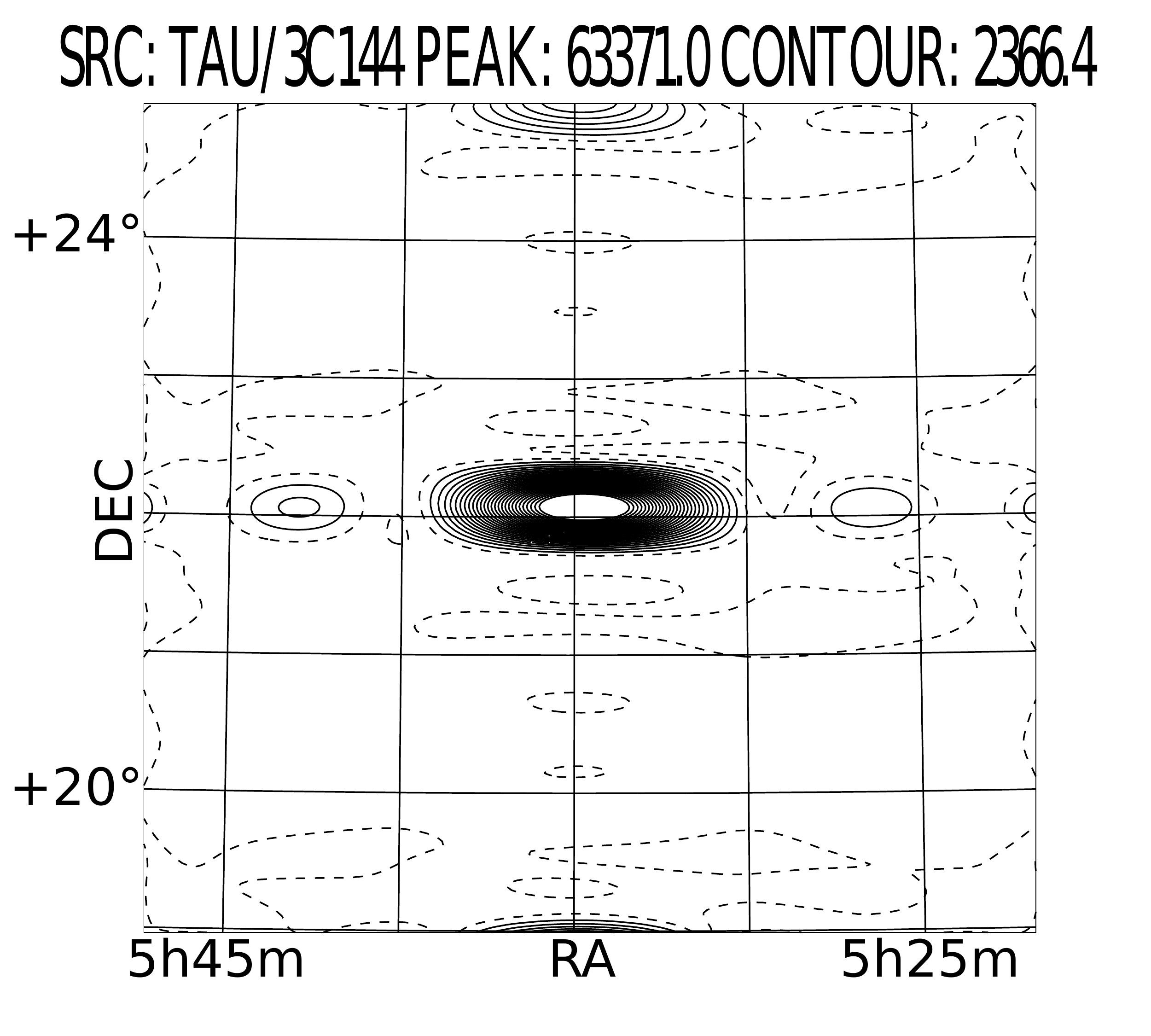}
        \label{fig:fs_cal}
    }
    \caption{
    Dirty images of Taurus A produced from the direct-imager visibilities. Figure \ref{fig:fs_uncal} is before complex gain corrections are applied during observations. Figure \ref{fig:fs_cal} is produced with complex gain corrections applied during observations.
    }\label{fig:fs_images}
\end{figure}

By applying gain solutions in the F-Engine during observations there is a drastic change in image quality as seen in Figure \ref{fig:fs_images}.
An otherwise distorted image of the Taurus A field becomes a bright point source.
The images in Figure \ref{fig:fs_images} have been obtained by first Fourier transforming the direct-imaging outputs into the \emph{uv} domain where they, and the equivalent FX correlator outputs, are imaged using traditional synthesis imaging software.

During the imaging process, the choice of weighting schemes introduces a small technical hurdle for direct-imaging correlator visibilities.
Since the effective baselines are an averaging of redundant baselines, using natural or uniform weighting in traditional imaging will produce equivalent results, that is a uniform-weighted image, with direct-imaging correlator visibilities.
In order to obtain the equivalent of a natural-weighted FX correlator image, the visibilities obtained by the direct-imaging correlator are given weights based on their redundancies.
In Figure \ref{fig:fs_images} we have used a natural-weighted scheme.

\subsubsection{System Stability}

\begin{figure*}
    \centering

    \subfigure[]{
        \includegraphics[width=.9\textwidth]{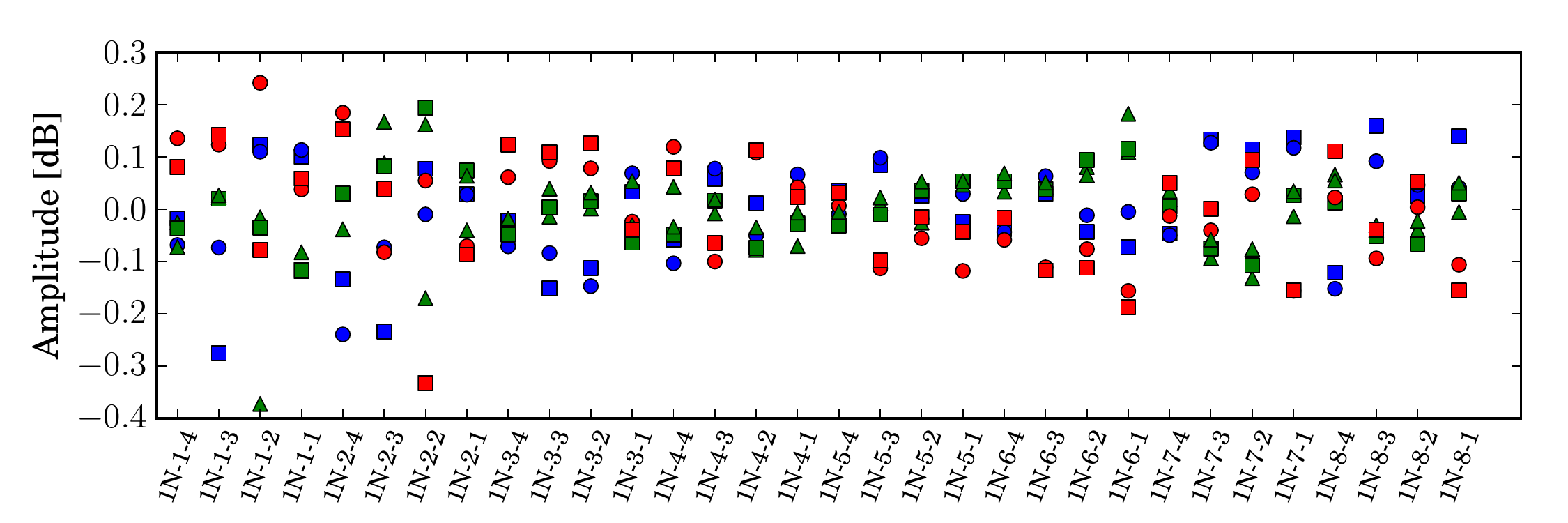}
        \label{fig:cal_amp_stability}
    }
    \vspace{-3mm}
    \subfigure[]{
        \includegraphics[width=.9\textwidth]{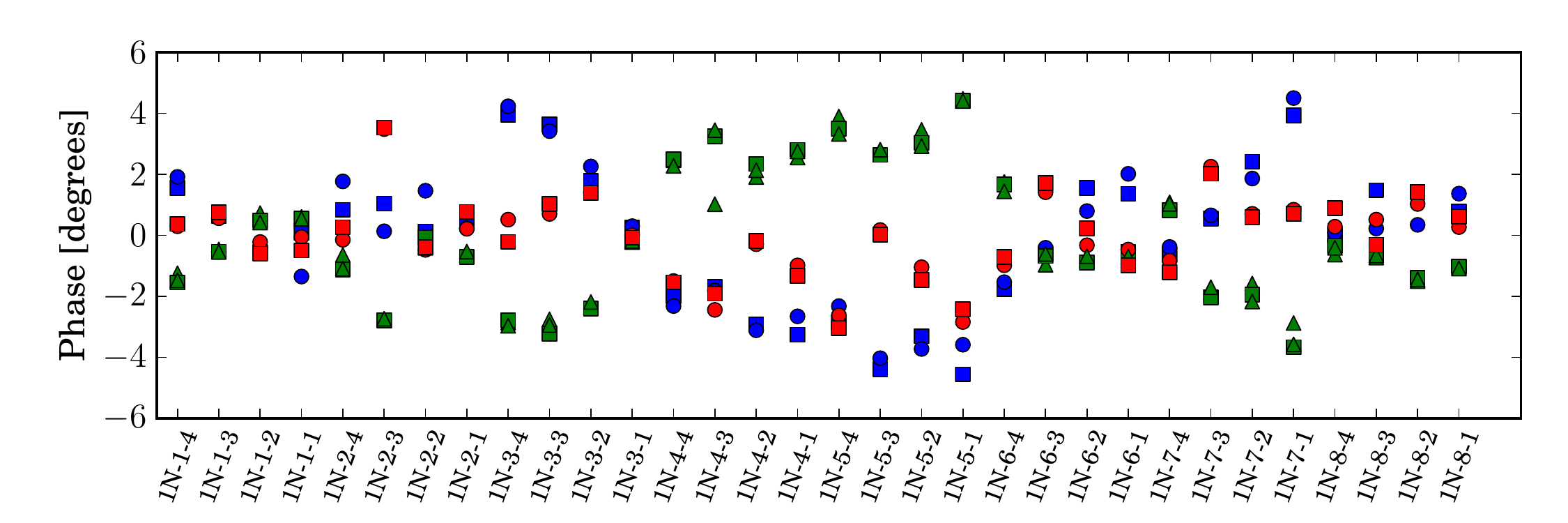}
        \label{fig:cal_phs_stability}
    }
    \caption{Complex gain solutions from 7 calibration observations using Cassiopeia A (blue), Taurus A (green), and Cygnus A (red).
    The amplitude is stable to within less than a dB over the two week period.
    The phase also remains stable, varying by only a few degrees in the same time period.
    There is an apparent solution bias based on the calibration source which is related to the array pointing angle.}
    \label{fig:cal_stability}
\end{figure*}

Since we have implemented a calibration method which uses a single bright point source as our sky model we need to consider the timescale on which the gain solutions are accurate.
Due to the transiting nature of \gls{best2} we are limited to only a few calibration opportunities per day, thus the calibration solutions need to be stable on at least multi-hour timescales.
Figure \ref{fig:cal_stability} shows the amplitude and phase solutions from 7 point source calibrations taken over a two week period for each of the receiving elements.
Over that period the amplitude and phase remain stable.
Calibration gains applied on day timescales has proven sufficient during our initial observations.

\subsection{Spatial FFT for Beamforming}

\begin{figure}
    \includegraphics[width=.46\textwidth]{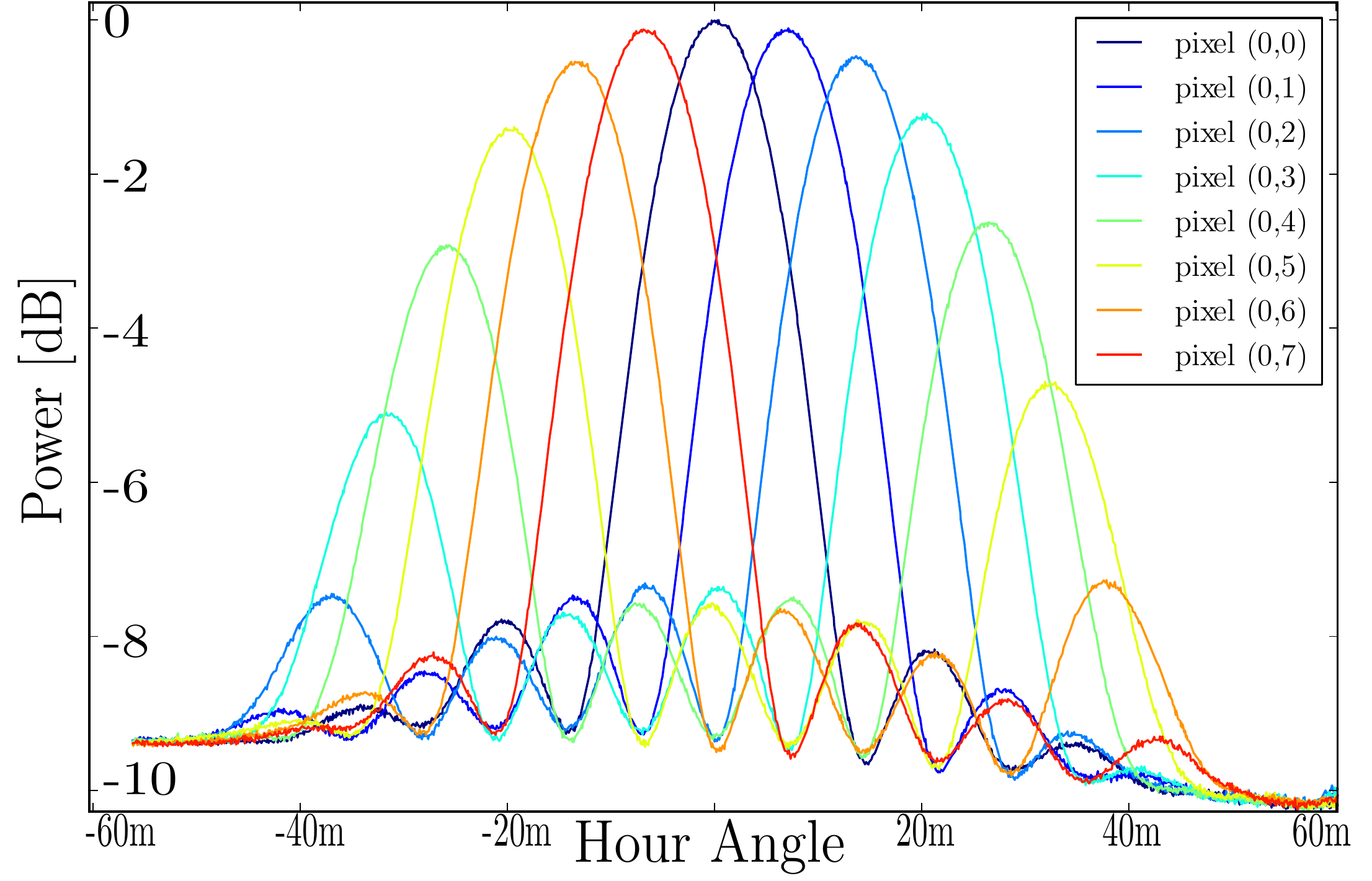}
    \caption{The normalized power observed in each of eight beams in an E-W array row during a transit of Cassiopeia A.
    The overall amplitude envelope is the primary beam response.}
    \label{fig:beam_transit}
\end{figure}

With the ability to calibrate antenna gains in real-time in place, the unaccumulated outputs of the 2-D \gls{fft} are calibrated electric-field beams on the sky. 
These beams remain fixed in position relative to the array location. 
Figure \ref{fig:beam_transit} shows the apparent flux of Cassiopeia A as it transits across the array.
Significant spatial aliasing effects are visible for all off-center beams, becoming more severe for beams further from the array phase centre. 
This is a result of the hierarchical, analogue beamformer used in \gls{best2}, the first layer of which is the summation of 16 critically spaced dipoles. 
Figure \ref{fig:beam_transit} also shows the loss of signal as the source moves between beams.

\begin{figure*}
    \includegraphics[width=.9\textwidth]{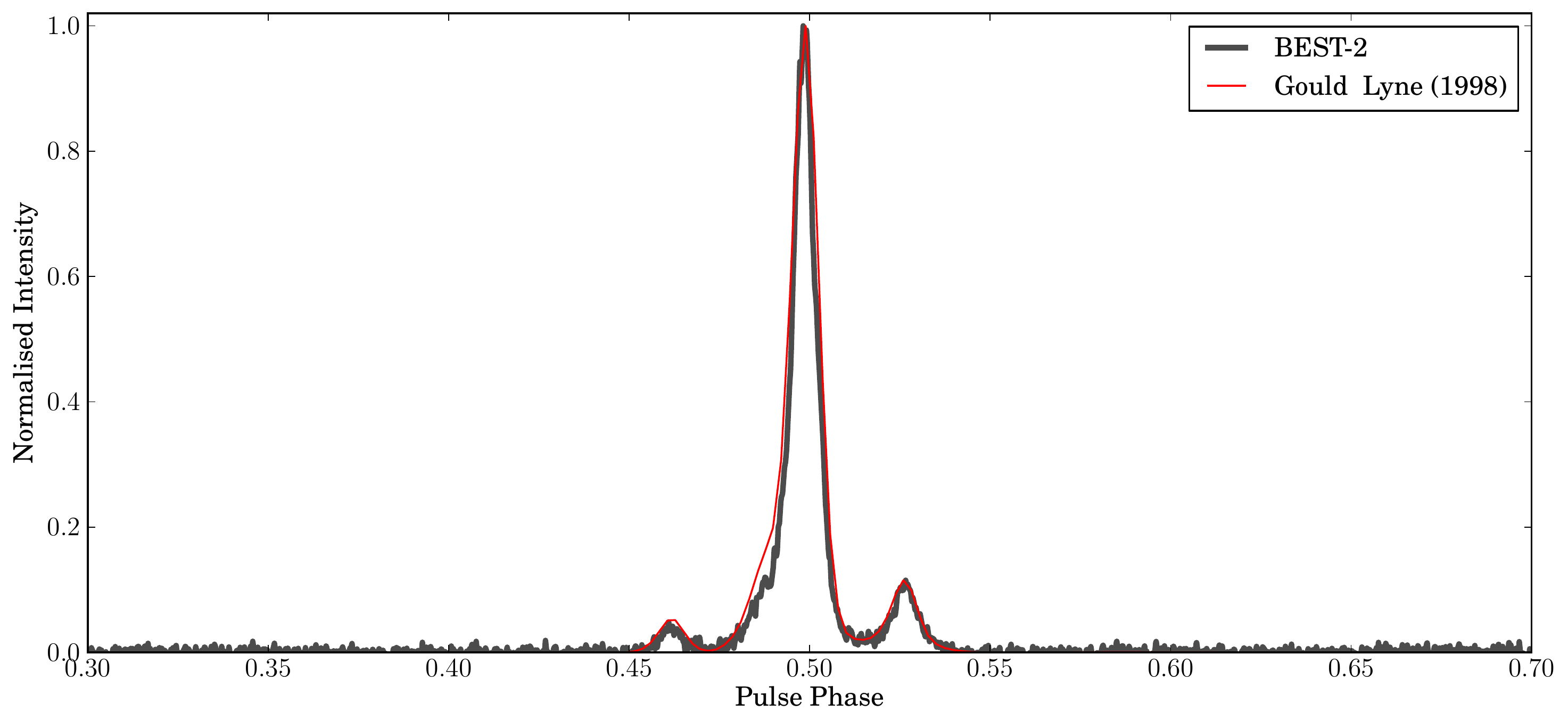}
    \caption{Folded profile (gray) of B0329+54 taken during an observation using BEST-2 with overplot (red) of 408 MHz Lovell Telescope archival data.}
    \label{fig:pulse_profile}
\end{figure*}

As noted earlier, this beamforming mode has already been put to use as the input to a pulsar and transient detection system \cite{multibeamgpu}.
This system was verified by detection of pulsar B0329+54.
After compensating for the DM of the pulsar, multiple pulses are summed to obtain a profile of the average pulse shape.
Figure \ref{fig:pulse_profile} shows a profile of the average pulse shape with a profile from 408 MHz Lovell Telescope archival data \cite{Gould21111998} overlaid.
With access to a complete set of beams which cover the field of view this system provides a promising platform for transient monitoring.
\section{Conclusion}

The main focus of this work has been on presenting a new digital backend for \gls{best2} and showing the currently available features in the design.
We have developed, deployed, and tested an \gls{fpga}-based direct-imaging correlator and beamformer along with an concurrent FX correlator system for the \gls{best2} array.
Additionally, we have developed the necessary software for general use of this instrument, it is currently in regular use at the observatory.
We have also touched on the data quality produced by the FX and direct-imaging correlators, and the beamformer.
We have demonstrated detection of various astronomical source with each of these instruments.
In a supplementary work we will present a study on the comparison of the FX and direct-imaging correlators.
That work will focus on the usability of the direct-imager for science observations, along with the challenge of calibrating such an instrument.

\section*{Acknowledgements}

We would like to thank the staff and support from Radiotelescopi di Medicina whom made this project possible, particularly Andrea Mattana, Frederico Perini, Germano Bianchi, Giovanni Naldi, Jader Monari, Marco Bartolini, Marco Schiaffino, and Stelio Montebugnoli.
Also, thank you to Ian Heywood for calibration and imaging help and advice.

\bibliography{refs}{}
\bibliographystyle{mn2e}

%
%

\section{Appendix: Instrumentation Block Diagrams}

This appendix is a supplement to "Implementation of a Direct-Imaging and FX Correlator for the BEST-2 Array".
Included are block diagrams of the digitizer/channelizer, FX correlator, and direct-imager components of the digital hardware.

\begin{figure*}
    \centering
\includegraphics[angle=90,scale=.4]{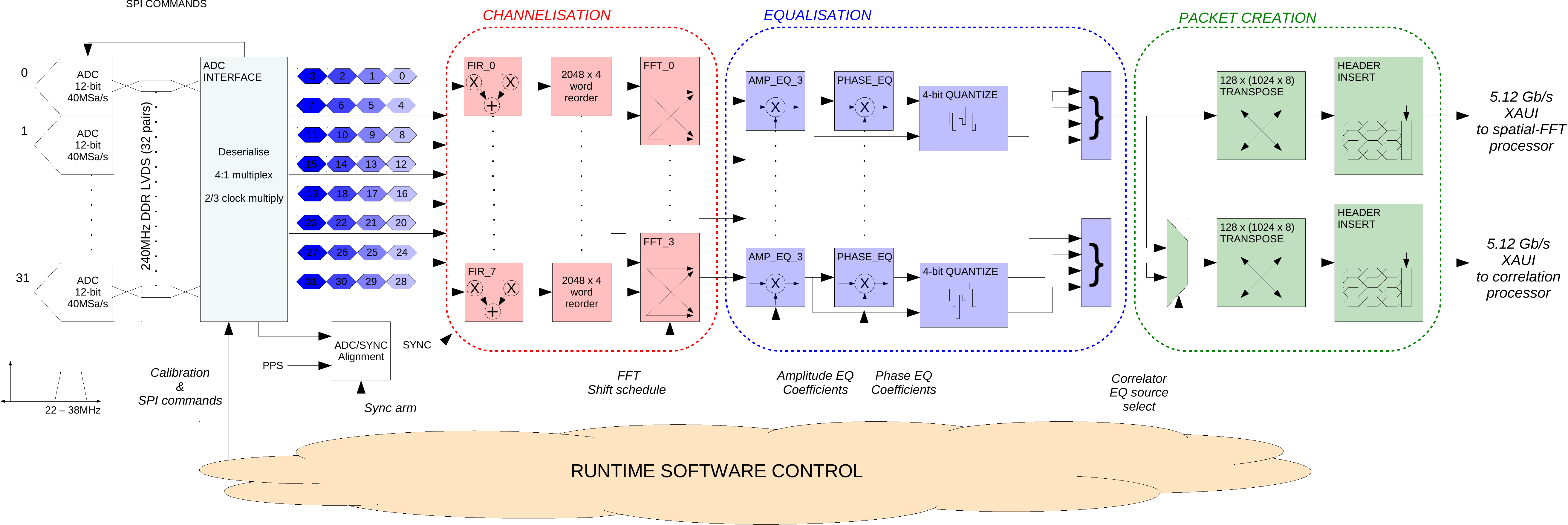}
    \caption{A block diagram describing the processing carried out on the ``F-engine''
ROACH board, which is used to channelize ADC data from an attached
daughterboard. Following channelization, software-defined amplitude
and phase corrections are applied to the frequency-domain data, which
is then quantized to 4-bit precision and output over XAUI for further
processing.}
    \label{fig:feng_block}
\end{figure*}

\begin{figure*}
    \centering
\includegraphics[angle=90,scale=.45]{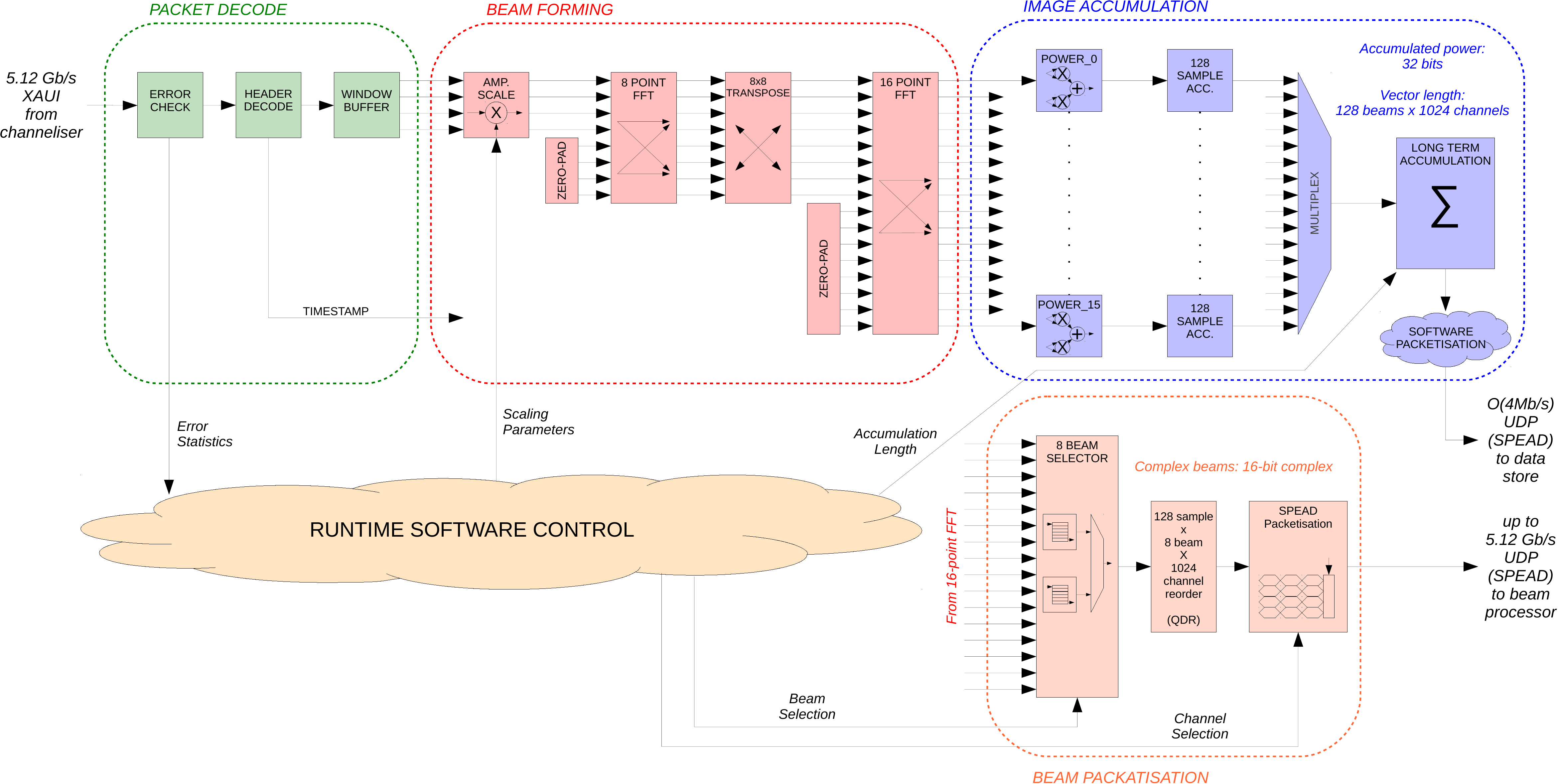}
    \caption{A block diagram describing the processing carried out on the ``S-engine'' 
ROACH board, which is used to perform a zero-padded spatial FFT on
an incoming data stream, forming a set of beams on the sky. Following
beamforming, accumulation of beam powers is performed to form low
time-resolution sky images. A runtime-programmable set of eight beams,
with user selectable frequency range are output via a 10GbE network,
for further processing of high time-resolution data.}
    \label{fig:seng_block}
\end{figure*}

\begin{figure*}
    \centering
\includegraphics[scale=.5]{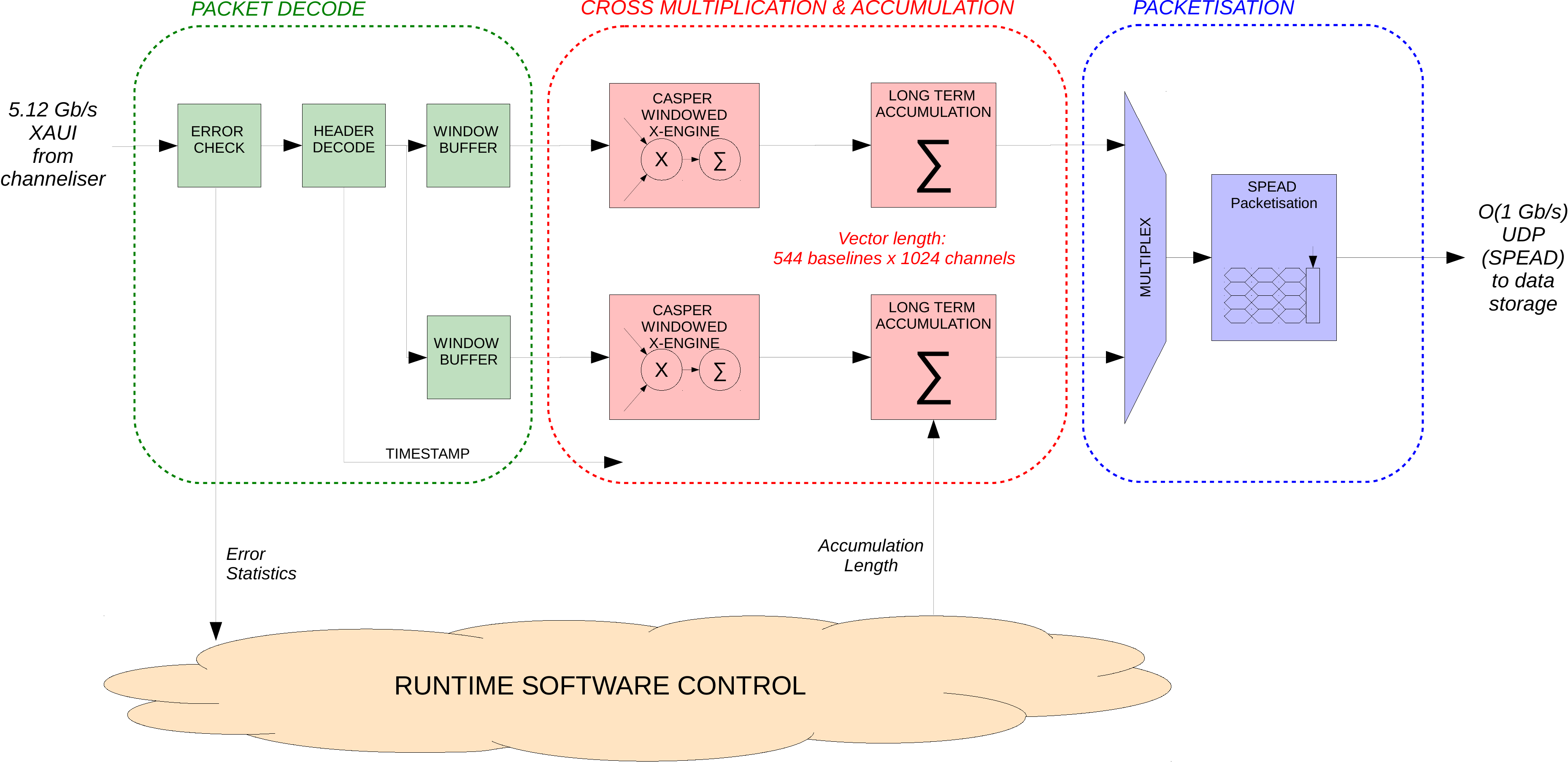}
    \caption{A block diagram describing the processing carried out on the ``X-engine'' 
ROACH board, which is used to perform cross multiplication and accumulation 
of antenna signals.}
    \label{fig:xeng_block}
\end{figure*}

\label{lastpage}


\end{document}